\documentclass{jfm}
\usepackage{graphicx}
\usepackage{epsfig} 
\usepackage{newtxtext}
\usepackage{newtxmath}
\graphicspath{{Figures/}}
\usepackage{multirow}
\usepackage{mathrsfs,amsmath}
\usepackage{bm}
\usepackage{amssymb}
\usepackage{multirow,bigdelim}
\usepackage{upgreek}
\usepackage{units}
\usepackage{array,multirow}
\usepackage{booktabs}
\usepackage{psfrag}
\usepackage{tikz}
\usepackage{caption}
\usetikzlibrary{arrows.meta}
\usepackage{color}
\usepackage{threeparttable}	
\usepackage{dcolumn}		
\newcolumntype{d}[1]{D{.}{.}{#1}}
\usetikzlibrary{decorations}
\usetikzlibrary{decorations.markings}
\usetikzlibrary{calc,decorations.pathreplacing}
\usepackage{lipsum}
\usepackage{mathtools}
\usepackage{xfrac}
\usepackage{ar}
\usepackage{enumitem}
\usepackage{url}
\usepackage{subcaption}
\usepackage{float}
\usepackage{tabularx}
\usepackage{natbib}
\usepackage{hyperref}
\usepackage{oplotsymbl}
\usepackage{soul}
\usepackage{xcolor}

\hypersetup{
    colorlinks = true,
    linkcolor = {blue},
    citecolor  = {black},
    filecolor = {blue},
    menucolor= {blue},
    runcolor = {blue},
    urlcolor = {blue},
}

\newcommand{\RomanNumeralCaps}[1]
\linenumbers

\definecolor{blueish}{rgb}{0, 0.44, 0.73}
\definecolor{darkred}{HTML}{A03B35}

\shorttitle{Wall-pressure--velocity correlations in high Reynolds number pipe flow}
\shortauthor{Dacome et al.}

\title{Scaling of wall-pressure--velocity correlations in high Reynolds number turbulent pipe flow}

\author{Giulio Dacome\aff{1} \corresp{\email{g.dacome@tudelft.nl, lorenzo.lazzarini7@unibo.it}}, Lorenzo Lazzarini\aff{2}\dag, Alessandro Talamelli\aff{2}, Gabriele Bellani\aff{2} and Woutijn J. Baars\aff{1}}
 
\affiliation{\aff{1}Faculty of Aerospace Engineering, Delft University of Technology, 2629 HS Delft, The Netherlands 
\aff{2}Department of Industrial Engineering / CIRI Aerospace, University of Bologna , 47121 Forl\`{i}, Italy}

\definecolor{blue1}{rgb}{0.81411765, 0.88392157, 0.94980392}
\definecolor{blue2}{rgb}{0.67189542, 0.81437908, 0.90065359}
\definecolor{blue3}{rgb}{0.47294118, 0.71163399, 0.85071895}
\definecolor{blue4}{rgb}{0.29098039, 0.5945098 , 0.78901961}
\definecolor{blue5}{rgb}{0.14666667, 0.46039216, 0.71869281}
\definecolor{blue6}{rgb}{0.0379085 , 0.32601307, 0.61830065}
\definecolor{blue7}{rgb}{0.03137255, 0.18823529, 0.41960784}

\begin{document}

\newcommand{\bluelineA}{\raisebox{2pt}{\tikz{\draw[-,blue1,solid,line width = 0.9pt](0,0) -- (5mm,0)}}}
\newcommand{\bluelineB}{\raisebox{2pt}{\tikz{\draw[-,blue2,solid,line width = 0.9pt](0,0) -- (5mm,0)}}}
\newcommand{\bluelineC}{\raisebox{2pt}{\tikz{\draw[-,blue3,solid,line width = 0.9pt](0,0) -- (5mm,0)}}}
\newcommand{\bluelineD}{\raisebox{2pt}{\tikz{\draw[-,blue4,solid,line width = 0.9pt](0,0) -- (5mm,0)}}}
\newcommand{\bluelineE}{\raisebox{2pt}{\tikz{\draw[-,blue5,solid,line width = 0.9pt](0,0) -- (5mm,0)}}}
\newcommand{\bluelineF}{\raisebox{2pt}{\tikz{\draw[-,blue6,solid,line width = 0.9pt](0,0) -- (5mm,0)}}}
\newcommand{\bluelineG}{\raisebox{2pt}{\tikz{\draw[-,blue7,solid,line width = 0.9pt](0,0) -- (5mm,0)}}}

\maketitle
\begin{abstract}
An experimental study was conducted in the CICLoPE long-pipe facility to investigate the correlation between wall-pressure and turbulent velocity fluctuations in the logarithmic region, at high friction Reynolds numbers ($4\,794 \lesssim Re_\tau \lesssim 47\,015$). Hereby we explore the scalability of employing wall-pressure to effectively estimate off-the-wall velocity states (\emph{e.g.}, to be of use in real-time control of wall-turbulence). Coherence spectra for wall-pressure and streamwise (or wall-normal) velocity fluctuations collapse when plotted against $\lambda_x/y$ and thus reveals a Reynolds-number-independent scaling with distance-from-the-wall. When the squared wall-pressure fluctuations are considered instead of the linear wall-pressure term, the coherence spectra for the wall-pressure--squared and velocity are higher in amplitude at wavelengths corresponding to large-scale streamwise velocity fluctuations (\emph{e.g.}, at $\lambda_x/y = 60$ the coherence value increases from roughly 0.1 up to 0.3). This higher coherence typifies a modulation effect, because low-frequency content is introduced when squaring the wall-pressure time series. Finally, quadratic stochastic estimation is employed to estimate turbulent velocity fluctuations from the wall-pressure time series only. For each $Re_\tau$ investigated, the estimated time series and a true temporal measurement of velocity inside the turbulent pipe flow, yield a normalized correlation coefficient of $\rho \approx 0.6$ for all cases. This suggests that wall-pressure sensing can be employed for meaningful estimation of off-the-wall velocity fluctuations, and thus for real-time control of energetic turbulent velocity fluctuations at high $Re_\tau$ applications.
\end{abstract}


\section{Introduction}\label{sec:intro}
Turbulence stresses in wall-bounded flows are inherently linked to the generation of skin-friction drag, and this prompts a significant interest in understanding their correlation with wall-based quantities \citep{renard:2016a}. In particular, the correlation between off-the-wall velocity fluctuations and wall-pressure fluctuations ($p_w$) is of significance in the context of using the latter as input to real-time flow control systems. That is, wall-based sensing requires the formulation of transfer functions \citep[\emph{e.g.},][]{sasaki:2019}, so that temporal dynamics of velocity structures can be inferred from non-intrusive wall-based measurements.

Studies on wall-pressure fluctuations of turbulent wall-bounded flows have focused on, amidst other aspects, the scaling of the pressure intensity and spectral signature. Scaling trends are a function of the friction Reynolds number, $Re_\tau \equiv \delta U_\tau/\nu$, where $\delta$ is the boundary layer thickness (in our work equal to the pipe radius, $R$), $U_\tau \equiv \sqrt{\tau_w/\rho}$ is the friction velocity (with $\tau_w$ being the wall-shear stress and $\rho$ being the fluid density) and $\nu$ is the fluid kinematic viscosity. Most notably, \citet{farabee:1991}, \citet{tsuji:2007} and \citet{klewicki:2008} revealed a characteristic inner-spectral peak in the wall-pressure spectra at a frequency of $f^+_p \approx 0.04$. Note that throughout the manuscript, quantities with a superscript `+' denote a normalization with the viscous length or time scale. The amplitude of said peak increases in magnitude with an increase in $Re_\tau$, as does the large-scale energy content. Efforts with direct numerical simulation (DNS) have confirmed these trends \citep[\emph{e.g.},][]{jimenez:2008,panton:2017,yu:2022a} and illustrated how, when considering spatial spectra, the inner-spectral peak resides at $\lambda_{x,p}^+ \approx 250$ (thus $f^+_p$ and $\lambda_{x,p}^+$ are related at the peak-scale through a streamwise convection velocity of $U_c^+ \approx 10$).

Relations between velocity structures and wall-pressure events have also been investigated. For instance, \citet{thomas:1983} revealed characteristic wall-pressure signatures associated with burst-sweep events in a turbulent boundary layer (TBL) flow, which are exclusively confined to the near-wall region. \citet{gibeau:2021} reported a low but significant scale-dependent coherence between wall-pressure, and streamwise ($u$) and wall-normal ($v$) velocity fluctuations in a TBL flow, at low frequencies (in the remainder of our manuscript lower-case quantities denote the fluctuations, while upper case ones signify time-averaged quantities). They ascribed this stochastic coupling to the passage of large-scale motions (LSMs). Recently, \citet{deshpande:2024a} assessed the growth of broadband energy in the wall-pressure spectrum by considering how the energy in velocity fluctuations, associated with active (producing turbulence kinetic energy) and inactive motions, scales with $Re_\tau$ and how this energy contributes to the energization of the intermediate and large pressure scales. Linking the wall-pressure field to the turbulence dynamics of LSMs is highly relevant for real-time flow control, because LSMs are a feasible target for an experimental implementation of control---because of their relatively long length- and time-scales---at application-level conditions of wall-turbulence \citep{abbassi:2017a,dacome:2024b}. At practically relevant values of $Re_\tau$, LSMs in the logarithmic region become energetically dominant over small scales \citep{hutchins:2007} and form the bulk of the turbulence kinetic energy production \citep{smits:2011b}. Moreover, a larger fraction of the turbulent velocity scales becomes strongly correlated across the wall-normal direction, and leaves a distinct imprint on the dynamics of near-wall turbulence and wall-pressure fluctuations \citep[\emph{e.g.},][]{marusic:2010,tsuji:2015a}.

Statistical analyses are typically adopted to quantify coherence aspects of velocity structures. It has been found that the minimum streamwise wavelength, $\lambda_{x,\text{min}}$, for which $u$ fluctuations are coherent between a location $y$ in the logarithmic region and a location in the near-wall region, follows $\lambda_{x,\text{min}}/y \approx 14$ (a scaling with the distance-from-the-wall). Furthermore, this scale threshold is invariant with $Re_\tau$ \citep{baars:2017a,baidya:2019a}. For the coherence between velocity fluctuations and \emph{wall-pressure fluctuations} instead, \citet{baars:2024} revealed a similar scaling but now with a scale threshold of $\lambda_{x,\text{min}}/y \approx 3$ (when considering $u$ fluctuations) and $\lambda_{x,\text{min}}/y \approx 1$ (when considering $v$ fluctuations). Again this scaling is invariant with $Re_\tau$, at least over the range of Reynolds numbers investigated with the DNS data of turbulent channel flow \citep[$Re_\tau \approx 550$ to 5\,200, from][]{lee:2015a}. It was also shown how the wall-pressure--squared signal contains a higher coherence with large-scale-filtered $u$ fluctuations, suggesting that the quadratic operator introduces large-scale energy content. This finding complied with an earlier conclusion of \cite{naguib:2001}, stating that the accuracy of stochastically estimating streamwise velocity fluctuations, from the unsteady wall-pressure, increases when incorporating a quadratic term.

The objective of our current work is to assess the scaling of the statistical correlation between wall-pressure and various components of the turbulent velocity in the logarithmic region of a fully-developed turbulent pipe flow. A unique experimental dataset was acquired with synchronised time series of wall-pressure and velocity, at $Re_\tau$ values ranging from ones that are typical of high-fidelity DNS, up to ones close to $Re_\tau = 50$k. We will address how the wall-pressure--velocity coherence adheres to a Reynolds-number-independent scaling for an unprecedented range of $Re_\tau$, and how current data compare to the ones available from the open literature. To this end, \S\,\ref{sec:methods} covers the experimental facility and measurement approach, and is followed by a description of the wall-pressure statistics in \S\,\ref{sec:pressure}. Subsequently, results for coherence of wall-pressure (and wall-pressure--squared) and streamwise velocity (\S\,\ref{sec:up}) and wall-normal velocity (\S\,\ref{sec:vwp}) are presented. Lastly, \S\,\ref{sec:est} builds upon the coherence results by analysing the accuracy of off-the-wall velocity estimates obtained solely from wall-pressure input data.

\section{Experimental methodology}\label{sec:methods}
\subsection{Experimental facility}
\begin{figure} 
\vspace{0pt}
\centering
\includegraphics[width = 0.999\textwidth]{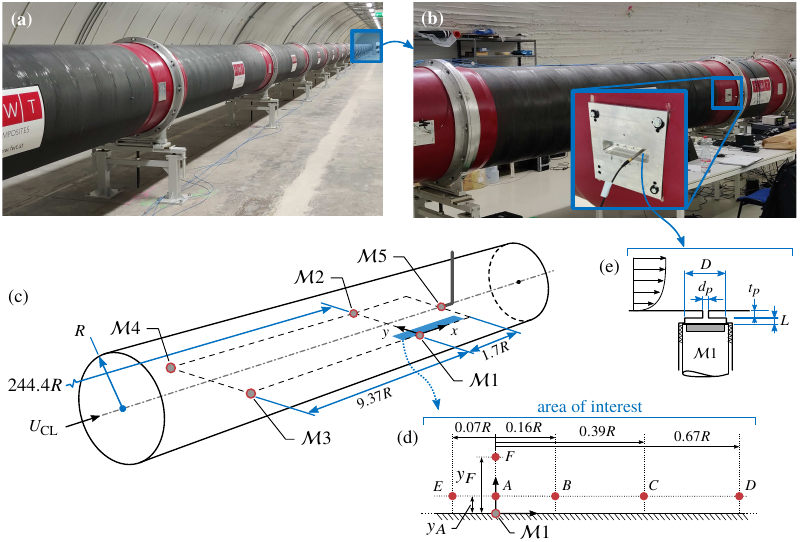}
\caption{(a,b) Photographs of the CICLoPE laboratory, with in (b) the test section at the downstream end of the long-pipe facility. (c) Schematic of the microphone sensor placement ($\mathcal{M}1$ to $\mathcal{M}4$ were mounted in the pipe wall, and $\mathcal{M}5$ was mounted along the pipe centreline). (d) Illustration of the points in the area of interest where acquisitions with single-wire and x-wire probes were performed. (e) Schematic of the pinhole--sub-surface-cavity, used to mount the microphones in the pipe wall.}
\label{fig:setup}
\end{figure}

An experimental campaign was carried out in the Centre for International Cooperation in Long-Pipe Experiments (CICLoPE, see Figs.~\ref{fig:setup}\textcolor{blue}{a,b}). The laboratory is realised inside a mountain to keep stable environmental conditions and to minimise background noise, while sound-absorbing material ensures minimal acoustic interference in the test section. The closed-loop facility comprises a 111.15\,m-long circular pipe with a radius of $R = D/2 = 0.4505$\,m. The primary streamwise location for measurements (where the flow is fully developed) is at $x^\prime = 110.1\,{\rm m} = 244.4R$ downstream of the pipe inlet. For the experiments reported, the pipe flow was operated at seven centreline velocities, with a maximum of $U_{\rm CL} = 44.60\,{\rm m/s}$ (corresponding to $Re_\tau \equiv U_\tau R/\nu = 47\,015$). Test conditions are elaborated upon in \S\,\ref{sec:resolution}. For presenting results, a Cartesian coordinate system is adopted with its origin at the primary streamwise location for measurements (at the centre of sensor $\mathcal{M}_1$, indicated in Fig.\,\ref{fig:setup}\textcolor{blue}{c}). Here the $x$-axis denotes the streamwise direction (positive in the downstream direction) and the $y$-axis denotes the wall-normal direction ($y = 0$ at the wall, and positive towards the centreline of the pipe). A comprehensive description of all design details of the facility can be found in the literature \citep{talamelli:2009a,bellani:2016}.

\subsection{Measurement instrumentation}\label{sec:instr}
Time-resolved pressure sensors were integrated in the wall of the pipe, each within its own cavity communicating with the flow through a pinhole orifice. Figure\,\ref{fig:setup}\textcolor{blue}{e} provides a schematic of the axisymmetric geometry of the pinhole and its corresponding sub-surface cavity, comprising a pinhole orifice diameter of $d_p = 0.3$\,mm, a pinhole depth of $t_p = 1.1$\,mm, an effective cavity diameter of $D = 4.6$\,mm and a cavity length of $L = 0.2$\,mm. The size of the pinhole orifice ensures a sufficient spatial measurement resolution for the purpose of the coherence analysis (\S\,\ref{sec:resolution}). However, because of the sub-surface-cavity geometry, a Kelvin-Helmholtz resonance occurs. This resonance phenomenon was quantified by means of an acoustic characterization experiment, following an identical procedure (in the exact same facility) as the one described by \citet[][pp.\,30-32]{baars:2024}. Similar procedures can be found in other works \citep[\emph{e.g.},][]{gravante:1998a,tsuji:2007,gibeau:2021}. In short, pressures at the orifice inlet ($p_i$) and within the cavity ($p_c$) were measured simultaneously, in quiescent flow conditions, under a broadband acoustic excitation in an anechoic facility. A linear transfer kernel was constructed, relating cavity to inlet pressure in the frequency domain: $H_r^{\rm exp}\left(f\right) = \langle \widetilde{P}_c(f) \widetilde{P}^*_i(f) \rangle / \langle  \vert \widetilde{P}_i(f) \vert^2 \rangle$. Here the angled brackets $\langle ... \rangle$ indicate ensemble averaging, the $*$ denotes the complex conjugate, and capitalised variables with a tilde indicate the Fourier transformed quantity, \emph{e.g.}, $\widetilde{P}_c(f) = \mathcal{F}\left[p_c(t)\right]$. Subsequently, a second-order model was fit to the gain of this transfer kernel and is denoted as $\vert H_r(f)\vert$. This procedure enabled the identification of the resonance frequency of the pinhole--sub-surface-cavity at $f_r = 4\,350$\,Hz. Implications of the resonance phenomenon on the wall-pressure measurements and coherence analyses are discussed later in \S\,\ref{sec:pressure}.

Regarding the pressure sensors themselves, GRAS\,46BE \nicefrac{1}{4}\,in. CCP free-field microphones were employed. These have an adequate dynamic range (35 to 160\,dB, with a reference pressure of $p_{\rm ref} = 20$\,$\mu$Pa) with an accuracy of $\pm 1$\,dB within the range of 10\,Hz to 40\,kHz. Data were acquired with two NI9234 analogue-input boards, comprising a 24-bit A/D conversion resolution. A total of five microphones were employed (labelled $\mathcal{M}1$ to $\mathcal{M}5$ in Fig.~\ref{fig:setup}\textcolor{blue}{c}): four were integrated in the pipe for wall-pressure measurements ($\mathcal{M}1$ to $\mathcal{M}4$), and one was mounted on a streamlined holder along the pipe centreline for monitoring the acoustic noise of the facility ($\mathcal{M}5$). Microphone $\mathcal{M}5$ was equipped with a GRAS\,RA0020 nose cone to reduce stagnation-driven turbulence pressure fluctuations on the otherwise flow-exposed diaphragm. The wall-mounted microphones were arranged in two streamwise pairs, separated by a distance of 4.22\,m ($\Delta x = 9.37R$). Microphones in one pair were located in azimuthally-opposite positions to facilitate the removal of facility (acoustic) noise.

Time series of streamwise velocity at two wall-normal locations in the logarithmic region ($y_{A} = 0.011$\,m $= 0.025R$ and $y_{F} = 0.061$\,m $= 0.135R$), and at five streamwise locations (points $A$ to $E$ in Fig.\,\ref{fig:setup}\textcolor{blue}{d}), were acquired using hot-wire anemometry (HWA). Synchronised measurements were performed of all microphones' signal at once, while velocity could only be measured at a single $y$-location for a given run. Each measurement was performed with an acquisition frequency of $f_s = 51.2$\,kHz, for an uninterrupted duration of $T_a = 480$\,s (relatively long time series were acquired to ensure sufficient convergence of the spectral statistics at the lowest frequencies of interest). A Dantec Streamline 90C10 CTA module was used, with a Dantec 55P15 single-wire boundary layer probe. Additionally, time series of the wall-normal velocity component were acquired using a Dantec 55P61 miniature x-wire probe at one point in the logarithmic region (point $A$ in Fig.\,\ref{fig:setup}\textcolor{blue}{d}). All Pt-plated tungsten wires of the single-wire and x-wire probes comprised sensing lengths of $l_{\rm hw}=1.25$\,mm and nominal diameters of $d_{\rm hw}=5\,\mu$m (thus, $l_{\rm hw}/d_{\rm hw} \approx 250$). Hot-wire probes were calibrated \emph{ex-situ} by employing a planar calibration jet. The single-wire probe was calibrated by fitting a 5$^{\rm th}$-order polynomial function to 11 calibration points of velocity versus measured voltage, $U = f(E)$. For the x-wire instead, seven velocity settings and thirteen angular positions were set to generate a two-dimensional look-up table \citep{burattini:2004} relating the two velocity components to the measured voltages of each wire: $\left(U_1,U_2\right) = f\left(E_1,E_2\right)$. During the measurements, the probe was oriented in such a way that it measured the streamwise ($u$) and wall-normal ($v$) velocity components simultaneously. More details of similar HWA measurements in the CICLoPE facility can be found in the works by \citet{orlu:2017a} and \citet{zheng:2022a}.

\subsection{Experimental conditions and measurement resolution}\label{sec:resolution}
\begin{table}
\centering
\begin{tabular}{lcccccccc}
 & Case & $\mathbf{1}$ & $\mathbf{2}$ & $\mathbf{3}$ & $\mathbf{4}$ & $\mathbf{5}$ & $\mathbf{6}$ & $\mathbf{7}$ \\
 & Label & \bluelineA & \bluelineB & \bluelineC & \bluelineD & \bluelineE & \bluelineF & \bluelineG \\
\multicolumn{9}{l}{}                                                                                                                                                                                                                     \\
\multirow{5}{*}{\rotatebox[origin=c]{90}{\textbf{\begin{tabular}[c]{@{}l@{}}Pipe flow\\ 
parameters\end{tabular}}}} & $Re_{\tau}$ & 4\,794 & 7\,148 & 14\,004 & 22\,877 & 31\,614 & 38\,271 & 47\,015 \\
 & $U_{\tau}$ (m/s) & 0.162 & 0.242 & 0.473 & 0.773 & 1.068 & 1.293 & 1.588 \\
 & $\tau_w$ (Pa) & 0.032 & 0.070 & 0.269 & 0.718 & 1.368 & 2.008 & 3.001 \\
 & $l^*$ ($\mu$m) & 94.0 & 63.0 & 32.2& 19.7 & 14.3 & 11.8 & 9.58 \\
 & $U_{\rm CL}$ (m/s) & 3.837 & 5.833 & 12.11 & 20.71 & 29.50 & 34.13 & 44.60 \\
\multicolumn{9}{l}{}                                                                                                                                                                                                                     \\
\multirow{9}{*}{\rotatebox[origin=c]{90}{\textbf{\begin{tabular}[c]{@{}l@{}}Instrumentation\\ 
characteristics\end{tabular}}}} & $d_p^+$ & 4.257 & 6.347 & 12.43 & 20.31 & 28.07 & 33.98 & 41.75 \\
 & $l^+_{\rm hw}$ & 13.30 & 19.84 & 38.86 & 63.48 & 87.72 & 106.2 & 130.5 \\
 & $f^+_{\rm s}$ & 29.71 & 13.36 & 3.482 & 1.305 & 0.683 & 0.466 & 0.301 \\
 & $T_aU_{\rm CL}/R$ & 4\,088 & 6\,214 & 12\,902 & 22\,066 & 31\,431 & 36\,364 & 47\,520 \\
\\
 & $y_A^+$ & 117.1 & 174.6 & 342.0 & 558.6 & 771.9 & 934.5 & 1\,148 \\    
 & $y_F^+$ & 649.2 & 968.0 & 1\,896 & 3\,097 & 4\,281 & 5\,182 & 6\,366 \\                  
 & \textit{\textbf{}} & & & & & & &  \\                                                         
\end{tabular}
\caption{Flow parameters corresponding to the seven test conditions in the CICLoPE long-pipe facility, with alongside nondimensional parameters of the instrumentation's geometry and acquisition details.}
\label{tab:parameters}
\end{table}
Seven experimental conditions were considered for measurements of the fluctuating wall-pressure and velocity in the CICLoPE long-pipe facility. Flow parameters of all test cases are reported in Table~\ref{tab:parameters}. With the aid of a heat exchanger, the facility was operated at constant temperature and the angular velocities of the two co-axial fans were set to generate centreline velocities in the range $3.837\,{\rm m/s} \leq U_{\rm CL} \leq 44.60\,{\rm m/s}$ (measured with a Pitot-static probe). Corresponding values of the wall-shear stress, $\tau_w$, were inferred from static pressure drop measurements \citep[following][]{fiorini:2017phd}. Values for the air density were indirectly measured with the air flow temperature and barometric pressure, so that values for the friction velocity, $U_\tau$, could be computed. For the experiments reported in this work, friction Reynolds numbers were in the range $4\,794 \lesssim Re_\tau \lesssim 47\,015$.

Spatial and temporal resolutions need to be considered for both the fluctuating wall-pressure and velocity measurements. For the measurement of wall-pressure, the pinhole orifice diameter dictates the spatial resolution, while for the measurement of velocity the hot-wire sensing length is determining the spatial resolution. The temporal resolution was limited by the acquisition frequency. All three parameters relevant for the measurement resolutions ($d_p$, $l_{hw}$ and $f_s$) are listed in Table~\ref{tab:parameters} after normalization with the viscous scales.

For fully-resolved wall-pressure measurements the pinhole orifice diameter must be $d_p^+ < 20$ \citep{gravante:1998a}. Hence, the pinhole diameter is not sufficiently small to claim  fully-resolved wall-pressure measurements for test cases 4 to 7 (the relatively large values of $d_p^+$ result in an attenuation of small-scale energy). However, this work does not revolve around conducting fully-resolved measurements, but rather focuses on the correlation between velocity fluctuations in the logarithmic region and wall-pressure. As reviewed in \S\,\ref{sec:intro}, the scales of interest for the correlation analyses reside at streamwise wavelengths of $\lambda_x/y \gtrsim 3$ (when considering $u$ fluctuations) and $\lambda_x/y \gtrsim 1$ (when considering $v$ fluctuations). Smaller streamwise scales in both the pressure and pressure-squared time series are not relevant, as they do not correlate with the ones in the turbulent velocity signals. Consequently, for all $Re_\tau$ test cases, a minimum streamwise wavelength that needs to be resolved for the coherence analyses is given by $\lambda_{x,\text{res}}/y_A = 1$ (recall that $y_A$ is the lowest wall-normal position being considered), resulting in a streamwise wavelength of $\lambda_{x,\text{res}} = y_A = 11$\,mm. The pinhole orifice diameter of $d_p = 0.3$\,mm is a factor 36.6 smaller and, thus, sufficient for capturing the streamwise wavelengths of interest.

When considering the spatial resolution of the HWA measurements, a similar reasoning can be applied. Statistically, the velocity structures of relevance to the wall-pressure--velocity correlations adhere to a self-similar scaling in all three dimensions. \citet{baidya:2019a} showed that the aspect ratio of coherent velocity structures is 7:1, in terms of their characteristic streamwise-to-spanwise length scales. Hence, the smallest structures of relevance have spanwise wavelengths of $\lambda_{z,\text{res}} = \lambda_{x,\text{res}}/7 \approx 1.6$\,mm. For the HWA measurements with the x-wire probe, the spanwise separation between both sensing wires is $\approx 1.0$\,mm, which is sufficient to resolve the scales of interest. In the $y$-direction the sensing length of the x-wire probe is also adequate, given the strong wall-normal coherence of the velocity structures. For the single-wire probe, its spanwise sensing length of $l_{\rm HW} = 1.25$\,mm is more than sufficient given that the $u$ fluctuations of interest are three times larger than the $v$ fluctuations of interest.

Regarding the temporal measurement resolution, this is set by the data acquisition rate. For the highest $Re_\tau$ test case (number 7), the acquisition time step is largest in terms of viscous time scales and equals $\Delta T^+ = 1/f_s^+ \approx 3.3$. Even though \citet{hutchins:2009} indicates a required time step $\Delta T^+$ of unity or less for fully-resolved measurements, the current acquisition rate is more than sufficient given the interest in much lower frequencies (larger spatial scales) than the ones corresponding to the dissipative regime.

\subsection{Post-processing of wall-pressure signals}\label{sec:filtering}
Even though the CICLoPE laboratory has been designed to minimise noise in the test section, the facility is non-anechoic and acoustic pressure fluctuations do contaminate the measured wall-pressure signals. A superposition of facility noise onto the time series of the hydrodynamic wall-pressure affects the wall-pressure statistics. Furthermore, the correlation analyses are affected since, by construction, facility noise and velocity fluctuations are uncorrelated. Therefore, a normalised correlation (with the additive facility noise present) is lower than the true value \citep{saccenti:2020}. 

Given the need to remove facility noise, a post-processing procedure is applied based on harmonic proper orthogonal decomposition \citep[hPOD, reviewed by][]{tinney:2020}. First, POD kernels are constructed from cross-spectral densities of, in this case, the various pressure signals. Then, the solution of an eigenvalue problem yields the frequency-dependent mode shapes and eigenvalues. By only retaining modes of the measured pressure time series, in which the spectral signature of facility noise is absent, hydrodynamic wall-pressure signals are inferred. All details of the noise-removal procedure are described in Appendix~\ref{app:noise}.

\section{Wall-pressure statistics in the CICLoPE facility}\label{sec:pressure}
Statistics of the wall-pressure fluctuations are presented to demonstrate the validity of our data for the correlation analyses presented in \S\,\ref{sec:up}-\ref{sec:vwp}.

\begin{figure} 
\vspace{0pt}
\centering
\includegraphics[width = 0.999\textwidth]{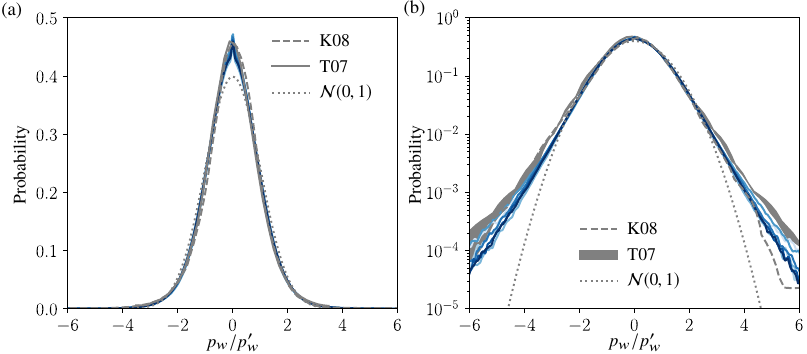}
\caption{Probability density functions of the wall-pressure fluctuations in the CICLoPE facility, for all $Re_\tau$ test cases considered (see Table~\ref{tab:parameters}). Current data are compared to a pdf obtained from atmospheric boundary layer data at $Re_\tau \approx 10^6$ \citep[][label K08]{klewicki:2008}, and a band representing the spread of pdf's obtained from zero-pressure-gradient TBL data at $1\,313 \lesssim Re_\tau \lesssim 3\,826$ \citep[][label T07]{tsuji:2007}. A standard $\mathcal{N}(0,1)$ Gaussian distribution is added for reference. Probability density functions are plotted with (a) a linear scale and (b) a logarithmic scale on the ordinate axes.}
\label{fig:pdf_pressure}
\end{figure}

Probability density functions (pdf's) of the wall-pressure time series are shown in Fig.\,\ref{fig:pdf_pressure}\textcolor{blue}{a} and Fig.\,\ref{fig:pdf_pressure}\textcolor{blue}{b}, with a linear and logarithmic scale on the ordinate axes, respectively. For both figures the amplitude-axes are scaled with the wall-pressure intensity (root-mean-square), denoted as $p^\prime_w$. Superimposed are several pdf's from the literature: a pdf corresponding to an atmospheric boundary layer flow at $Re_\tau \approx 1 \times10^6 \pm 2\times 10^5$ \citep{klewicki:2008}, and a band representing the spread of pdf's corresponding to zero-pressure-gradient TBL flow at $1\,313 \lesssim Re_\tau \lesssim 3\,826$ \citep{tsuji:2007}. All pdf's of the current datasets show negligible disparity between the test cases, and are consistent with the distributions from the literature. Minor deviations appear in the tails of the pdf's (Fig.\,\ref{fig:pdf_pressure}\textcolor{blue}{b}), yet comparable with the degree of deviation in the work by \citet{tsuji:2007} and without a noticeable monotonic trend with an increase in $Re_\tau$.

\begin{figure} 
\vspace{0pt}
\centering
\includegraphics[width = 0.999\textwidth]{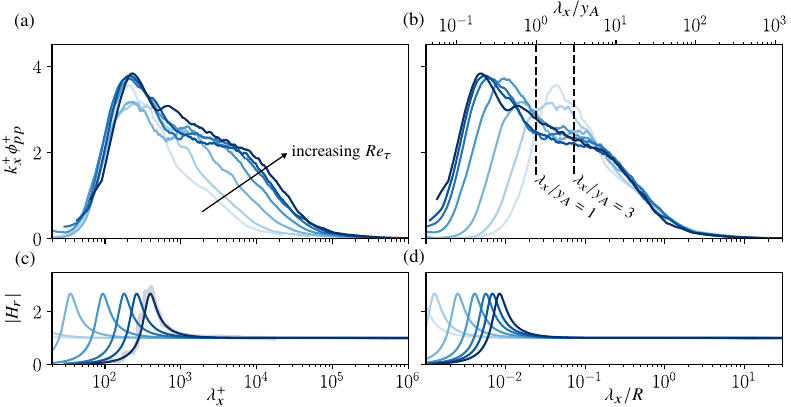}
\caption{(a,b) Pre-multiplied energy spectra of wall-pressure fluctuations, for all $Re_\tau$ test cases considered (an increase in colour intensity of corresponds to an increase in $Re_\tau$ following test cases $1\rightarrow7$, listed in Table~\ref{tab:parameters}), as a function of (a) the viscous scaled wavelength and (b) the outer-scaled wavelength. Note that the temporal spectra are plotted as spatial spectra by converting frequency into wavelength, using $\lambda_x \equiv U_c/f$ with $U_c^+ = 10$. Vertical dashed lines in sub-figure (b) indicate the minimum wavelength in our dataset for which wall-pressure--velocity correlations become appreciable when considering $u$ fluctuations ($\lambda_x/y_A = 3$) and $v$ fluctuations ($\lambda_x/y_A = 1$). (c,d) Gain of transfer kernel $H_r$ that characterizes the pinhole--sub-surface-cavity as described in \S\,\ref{sec:instr}, including in (c) the gain of the raw kernel, $H_r^{\rm exp}$ (light grey line).}
\label{fig:spectra_pressure}
\end{figure}

Pre-multiplied energy spectra of wall-pressure fluctuations are shown in Fig.\,\ref{fig:spectra_pressure}\textcolor{blue}{a} and Fig.\,\ref{fig:spectra_pressure}\textcolor{blue}{b} for all values of $Re_\tau$, with an inner-scaled and outer-scaled streamwise wavelength on the abscissa, respectively. Here, the streamwise wavelength is obtained by applying Taylor's hypothesis: $\lambda_x \equiv U_c/f$, where $f$ is the frequency, and $U_c$ is the convection velocity taken as $U_c^+ = 10$. Despite the convection velocity of the wall-pressure field being scale-dependent \citep[\emph{e.g.},][]{luhar:2014}, and the temporal-to-spatial conversion of near-wall fluctuations in velocity/pressure not abiding by Taylor's hypothesis \citep{dennis:2008a,delalamo:2009}, the conversion is kept equal across all test cases. In essence, we compare temporal spectra (since $\lambda_x^+ = 10U^2_\tau/\nu/f = 10/f^+$). Still the temporal-to-spatial conversion was applied because \S\,\ref{sec:up}--\ref{sec:vwp} consider all coherence spectra as a function of wavelength for ease of comparison to the only data available (those from \emph{spatial} DNS of turbulent channel flow).

Before commenting on the wall-pressure spectra, note that Fig.\,\ref{fig:spectra_pressure}\textcolor{blue}{c} and Fig.\,\ref{fig:spectra_pressure}\textcolor{blue}{d} present the gain of the transfer kernel that characterizes the pinhole--sub-surface-cavity (described in \S\,\ref{sec:instr}). Because the transfer kernel is a function of frequency, and the frequency-to-wavelength conversion includes the friction velocity of each test case, seven identical kernels are shown (but shifted along $\lambda_x$). For reference, the raw experimental kernel, $H_r^{\rm exp}$, is shown in Fig.\,\ref{fig:spectra_pressure}\textcolor{blue}{c} for the highest $Re_\tau$ test case with a thick grey line, while the other kernels correspond to the fitted kernel of the second-order model, $H_r$. Noticeably, resonance occurs at scales where the wall-pressure spectra are energetic. It is thus necessary to correct the spectra for the amplification/attenuation effect. Current wall-pressure spectra \emph{were corrected} before plotting, by dividing the spectra with the frequency-dependent model kernel [$\phi_{pp}(f) = \phi_{pp,f}(f)/\vert H_r(f)\vert^2$, with $\phi_{pp,f}$ being the spectrum after removing facility noise from the raw measurements of wall-pressure following Appendix~\ref{app:noise}]. The resonance-correction works theoretically, but practically the kernel (which was found with the aid of a flow-off experiment) changes when wall-bounded turbulence grazes the pinhole orifice \citep[see][]{dacome:2024c}, making the correction imperfect. In practice, this results in an erroneous `wiggle' in various spectra, and is most noticeable in the spectrum of test case 7.

Close inspection of the wall-pressure spectra reveals expected Reynolds-number trends. At first, the location of the inner-spectral peak at $\lambda_{x,p}^+ \approx 250$ (Fig.\,\ref{fig:spectra_pressure}\textcolor{blue}{a}) agrees well with previous findings \citep{farabee:1991,tsuji:2007,klewicki:2008,panton:2017}. A slight increase in the inner-spectral peak magnitude, with an increase in $Re_\tau$, is also noticeable for test cases 4 to 7 \citep[expected per the trends in][]{tsuji:2007,panton:2017,yu:2022a}. The large-scale energy content also progressively increases with $Re_\tau$ and exhibits a collapse in outer-scaling, for the range $\lambda_x/R \gtrsim 0.2$ (Fig.\,\ref{fig:spectra_pressure}\textcolor{blue}{b}). This trend is in line with the findings reported in DNS studies at lower $Re_\tau$ \citep{panton:2017,yu:2022a}. It also conforms to the work by \citet{deshpande:2024a}, who reason that the intermediate and large scales of the wall-pressure spectra grow with $Re_\tau$ due to the contributions of both the active and inactive motions in the grazing flow. Spectra corresponding to test cases 1 and 2 are outliers in that their broadband peak-magnitudes are relatively high. We postulate that this is due to an incomplete removal of facility noise, as any remaining signature of facility noise is more pronounced in the spectra of lower $Re_\tau$ test cases. That is, the degree of facility noise was quantified with a signal-to-noise ratio (SNR), defined as the intensity-ratio of turbulence-induced wall-pressure fluctuations, relative to those induced by facility noise: ${\rm SNR} = p^\prime_w/(p^\prime_{w,r} - p^\prime_w)$. Here, $p^\prime_{w,r}$ is the pressure intensity (root-mean-square) of the raw, measured wall-pressure. SNRs in our dataset increase monotonically with $Re_\tau$, in the interval $0.08 \leq {\rm SNR} \leq 0.25$. Additive facility noise is thus more noticeable in the spectra at lower $Re_\tau$. For the remainder of the paper, it is important to recall from \S\,\ref{sec:resolution} that for the correlation analysis the scales of interest reside at streamwise wavelengths beyond $\lambda_x/y \approx 3$ (when considering $u$ fluctuations) and $\lambda_x/y \approx 1$ (when considering $v$ fluctuations). Both of these limits are indicated in Fig.\,\ref{fig:spectra_pressure}\textcolor{blue}{b}; within the scale-range of interest the spectra are not affected by the kernel-correction and only the two lowest test cases seem affected by additive (acoustics-driven) noise.

\begin{figure} 
\vspace{0pt}
\centering
\includegraphics[width = 0.999\textwidth]{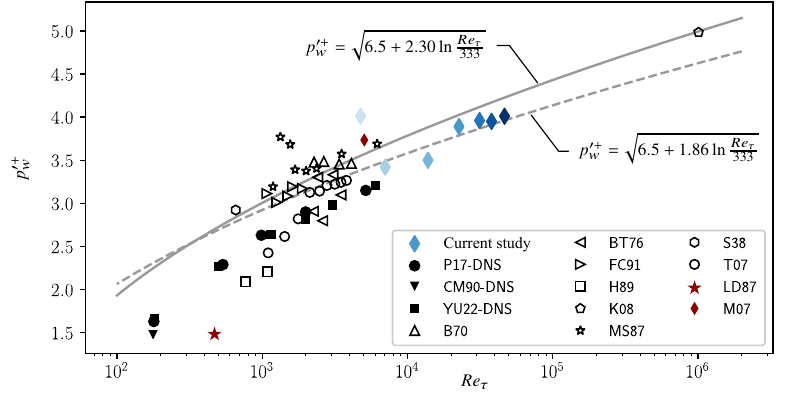}
\caption{Wall-pressure intensity inferred from integrating the wall-pressure spectra. Current results are compared to several datasets available from the literature. Data are taken from the DNS studies of \cite{panton:2017} (P17-DNS, $\bullet$, ZPG-TBL), \cite{choi:1990} (CM90-DNS, $\blacktriangledown$, TCF) and \cite{yu:2022a} (YU-DNS, $\blacksquare$, pipe flow). Furthermore, data are collected from experimental studies of ZPG-TBL flows: \cite{blake:1970} (B70, $\triangle$), \cite{bull:1976} (BT76, $\triangleleft$), \cite{farabee:1991} (FC91, $\triangleright$), \cite{horne:1989} (H89, $\square$), \cite{klewicki:2008} (K08, $\pentago$), \cite{mcgrath:1987} (MS87, $\starlet$), \cite{schewe:1983} (S83, $\hexago$) and \cite{tsuji:2007} (T07, $\circ$), and of experimental studies of pipe flows: \cite{lauchle:1987} (LD87, \textcolor{darkred}{$\bigstar$}) and \cite{morrison:2007} (M07, \textcolor{darkred}{$\blacklozenge$}). Solid and dashed lines are the formulations presented by \citet{klewicki:2008}, in which the pressure variance increases logarithmically with increasing $Re_\tau$.}
\label{fig:intensity}
\end{figure}

As a final wall-pressure statistic, we consider the wall-pressure intensities, resulting from the integration of the energy spectra. Here, the root-mean-square intensity is considered and inner-normalized following $p^{\prime +}_w = p^{\prime}_w/\tau_w$. Wall-pressure intensities are plotted in Fig.\,\ref{fig:intensity} and compared to a variety of datasets from the literature. Data from channel flow DNS are added \cite{panton:2017}, together with the various datasets assembled by \cite{klewicki:2008} (and named in the caption), that include both numerical and experimental studies, comprising zero-pressure-gradient turbulent boundary layer (ZPG-TBL), turbulent channel (TCF) and pipe flows. Our current data confirms the trend of increasing pressure intensity with $Re_\tau$, and closely follows the empirical relation of \cite{klewicki:2008}. Only the data point of test case 1 (at $Re_\tau \approx 4\,794$) is an outlier, which is ascribed to the imperfect facility noise-filtering causing an overestimation of the wall-pressure intensity.

\section{Coherence between streamwise velocity and wall-pressure fluctuations}\label{sec:up}
To analyse the scale-dependent coupling between the fluctuations in streamwise velocity ($u$) and wall-pressure ($p_w$), the linear coherence spectrum (LCS) is employed. 
The LCS describes the stochastic coupling, on a per-scale basis, as the degree of \emph{phase-consistency}. The LCS is defined as the magnitude-squared of the cross-spectrum between $u$ and $p_w$, normalized with the two auto-spectra of $u$ and $p_w$:
\begin{equation}\label{eq:LCS}
    \gamma_{up_w}^2\left(y,\lambda_x\right) \equiv \frac{\vert \langle  \widetilde{U}\left(y,\lambda_x\right) \widetilde{P}^*_w\left(\lambda_x\right) \rangle \vert^2}{\langle \vert \widetilde{U}\left(y,\lambda_x\right) \vert^2 \rangle \langle \vert \widetilde{P}_w\left(\lambda_x\right)\vert^2 \rangle},
\end{equation}
where the angled brackets $\langle ... \rangle$ indicate ensemble averaging, the $*$ denotes the complex conjugate, and capitalised variables with a tilde indicate the Fourier transformed quantity, \emph{e.g.}, $\widetilde{P}_w(f) = \mathcal{F}\left[p_w(t)\right]$. Because in the remainder of the manuscript we present scale-dependent data as a function of streamwise wavelength, the argument in \eqref{eq:LCS} is taken as $\lambda_x$ and is, as for the energy spectra in \S\,\ref{sec:pressure}, obtained by applying Taylor's hypothesis: $\lambda_x \equiv U_c/f$, with $U_c^+ = 10$.

\begin{figure} 
\vspace{0pt}
\centering
\includegraphics[width = 0.999\textwidth]{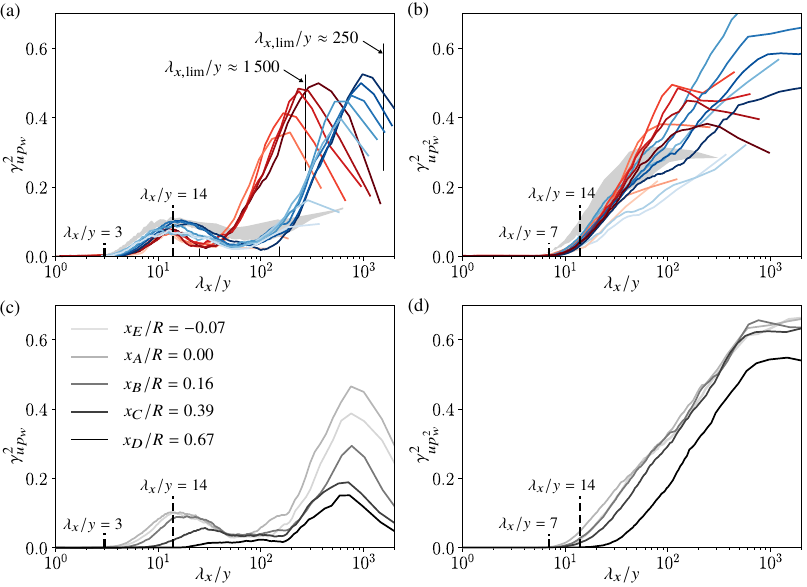}
\caption{(a) Coherence spectra for the fluctuations in streamwise velocity and wall-pressure, and (b) the streamwise velocity and wall-pressure--squared. Two sets of coherence spectra are shown, corresponding to velocity fluctuations measured at point $A$ (blue colour scale) and point $F$ (red colour scale); an increase in colour intensity corresponds to an increase in $Re_\tau$ following test cases $1\rightarrow7$, listed in Table~\ref{tab:parameters}. Reference data are shown with a light grey shaded area, associated with the spread of coherence spectra from spatial DNS data at $Re_\tau = 5\,200$ \citep{baars:2024}. (c) Coherence spectra for the fluctuations in streamwise velocity and wall-pressure, and (d) the streamwise velocity and wall-pressure--squared, for test case 3 ($Re_\tau \approx 14\,004$), and for velocity fluctuations measured at points E, A-D spanning a range of streamwise locations, $-0.07 \leq x/R \leq 0.67$. Note that all current coherence spectra are generated from temporal data, and plotted as spatial spectra by converting frequency into wavelength using $\lambda_x \equiv U_c/f$ with $U_c^+ = 10$.}
\label{fig:up_coupling}
\end{figure}

Figures\,\ref{fig:up_coupling}\textcolor{blue}{a,b} present the LCS for $u$ and $p_w$, for two positions of the velocity measurement (points $A$ and $F$ in Fig.\,\ref{fig:setup}\textcolor{blue}{d}) and for all values of $Re_\tau$. In presenting the scale-dependent spectra we resort to scaling $\lambda_x$ with the distance-from-the-wall, so that the abscissae are in terms of $\lambda_x/y$ with $y = \{y_{A},y_{F}\}$ for these graphs. With negligible coherence reported at small wavelengths, a steady rise in the LCS can be observed in Fig.\,\ref{fig:up_coupling}\textcolor{blue}{a} for increasing $\lambda_x/y$ until a local maximum is reached at $(\lambda_x/y,\gamma_{up_w}^2) \approx (14,0.1)$. Given that the current data span nearly a decade in $Re_\tau$ we can conclude that the region of coherence centred at $\lambda_x/y \approx 14$ is Reynolds-number invariant. Only the two LCS corresponding to test cases 1 and 2 have a slightly lower value near $\lambda_x/y = 14$. This is ascribed to the incomplete removal of facility noise in the wall-pressure spectra of these two test cases (recall the discussion in \S\,\ref{sec:pressure}), and the fact that additive noise in the spectra causes an attenuation of the LCS. A second region of significant coherence appears at large wavelengths where the velocity fluctuations continue to be energetically relevant. To illustrate this, an amplitude threshold of the pre-multiplied streamwise energy spectra is taken as $k^+_x\phi^+_{uu} = 0.2$ at the large-scale end. Energy levels only drop below this threshold for outer-scaled wavelengths of $\lambda_{x,\mathrm{lim}}/R \gtrsim 35$. This limit is included in Fig.\,\ref{fig:up_coupling}\textcolor{blue}{a} for reference, and corresponds to $\lambda_{x,\mathrm{lim}}/y \approx \{250,1\,500\}$ for $y = \{y_A,y_F\}$, respectively. Moreover, this region of high coherence (in excess of $\gamma^2_{up_w} = 0.5$) becomes relevant at a scale of $\lambda_{x,\mathrm{lim}}/R \approx 3.5$ (or $\lambda_{x,\mathrm{lim}}/y \approx \{25,150\}$ for $y = \{y_A,y_F\}$ as indicated with thin solid lines in Fig.\,\ref{fig:up_coupling}\textcolor{blue}{a}), and thus adheres to a Reynolds-number invariant scaling when $\lambda_x$ is scaled with the outer-scale, $R$. This large-scale region of strong coherence between $u$ fluctuations in the logarithmic region and the wall-pressure field is presumably related to global velocity modes \citep{bullock:1978a,delalamo:2003a}. These global modes are `inactive' in the view of Townsend's attached-eddy hypothesis \citep{townsend:1976bk} (thus large-scale eddies that do not contribute to the Reynolds shear stress $uv$). Inactive motions are coupled to the very large scales in the pressure spectrum \citep[as shown explicitly by][]{deshpande:2024a}, while the active motions contribute directly to the intermediate scales. 

Before further discussing the trends of the coherence spectra, we proceed with inspecting the coherence involving the quadratic term of the wall-pressure. The inclusion of this term was deemed important for stochastically estimating off-the-wall velocities from wall-pressure data. The quadratic term of the wall-pressure is taken $p_w^2 = \left[p_w^2\right]_{\rm r} - \overline{\left[p_w^2\right]_{\rm r}}$ with $\left[p_w^2\right]_{\rm r}$ denoting the time series of the wall-pressure--squared prior to the subtraction of its mean. Differently to the behaviour displayed by the linear term of wall-pressure, the LCS for $u$ and $p_w^2$ rises starting from $\lambda_x/y \approx 7$ (see Fig.\,\ref{fig:up_coupling}\textcolor{blue}{b}). Again a Reynolds-number invariant trend appears in the rise of coherence around scales of $\lambda_x/y = 14$ and beyond, with once more the LCS of test cases 1 and 2 comprising a lower magnitude due to the incomplete removal of facility noise from the wall-pressure spectra. To further conclude the Reynolds-number invariant trends observed in Figs.\,\ref{fig:up_coupling}\textcolor{blue}{a,b}, the current experimental coherence spectra generated from temporal data are compared to the coherence spectra presented by \citet{baars:2024}, generated from spatial DNS data of turbulent channel flow. These reference data are shown with the light grey shaded area, indicating the spread of coherence spectra at $Re_\tau = 5\,200$ when considering a range of wall-normal positions across the logarithmic region \citep[$80 \lesssim y^+ \lesssim 0.15Re_\tau$, see Fig.\,6 of][]{baars:2024}. Moreover, \citet{baars:2024} also revealed a Reynolds-number invariant trend for these DNS data, spanning $550\lesssim Re_\tau \lesssim \approx 5\,200$. It must be noted that even though these DNS data are associated with turbulent channel flow, it was shown that coherence spectra from a relatively low Reynolds number TBL flow ($Re_\tau \approx 2\,280$) were also in agreement with these channel flow data. And so, with the current LCS for pipe flow collapsing for the full range of $Re_\tau$, for both wall-normal positions, $y = \{y_{A},y_{F}\}$ (while agreeing with the reference data), it can be concluded that the coherence is statistically similar across several canonical flow geometries.

Universal trends in the coherence spectra are reflective of how coherent velocity fluctuations are interlinked to the wall-pressure. The scales around which $\gamma^2_{up_w}$ and $\gamma^2_{up^2_w}$ become non-zero ($\lambda_x/y \approx 3$ and $\lambda_x/y \approx 7$, respectively), as well as the logarithmic growth of coherence [most noticeable in Fig.\,\ref{fig:up_coupling}\textcolor{blue}{b}, where $\gamma^2_{up^2_w} \propto \ln\left(\lambda_x/y\right)$] follow a pattern presented in the work by \citet{baars:2017a}. They considered the coherence between the near-wall velocity fluctuations and the ones in the logarithmic region. The logarithmic growth of coherence, which occurs over the inertial-range of wavelengths, was interpreted as the range of scales that contains turbulence energy that is statistically self-similar (following a hierarchical structure of wall-attached eddies).

\begin{figure} 
\vspace{0pt}
\centering
\includegraphics[width = 0.999\textwidth]{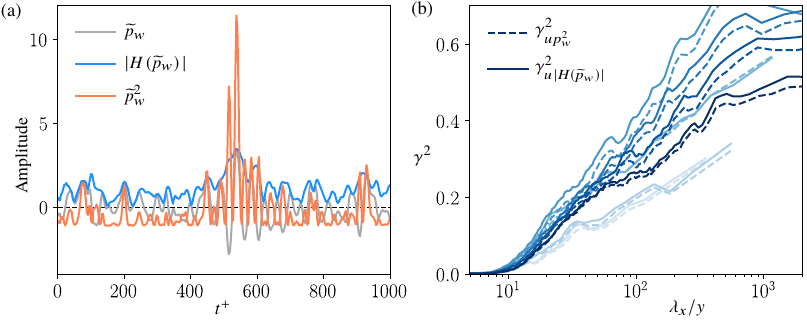}
\caption{(a) Normalised wall-pressure time series of microphone $\mathcal{M}1$, for test case 3 ($Re_\tau \approx 14\,004$), its Hilbert transform, and the corresponding de-meaned wall-pressure--squared. (b) Coherence spectra for the fluctuations in streamwise velocity and wall-pressure--squared (dashed lines, identical to the coherence spectra in Fig.\,\ref{fig:up_coupling}\textcolor{blue}{b}), compared to the coherence spectra for fluctuations in the streamwise velocity at point $y_A$ and the Hilbert transform of the wall-pressure (solid lines).}
\label{fig:hilbert}
\end{figure}

An increase in large-scale coherence for the wall-pressure--squared term suggests that large-scale $u$ fluctuations modify (modulate) the wall-pressure field following nonlinear dynamics. To analyse this phenomenon, a Hilbert transform is used to retrieve an ``envelope" of the wall-pressure time series. Figure\,\ref{fig:hilbert}\textcolor{blue}{a} presents the normalised wall-pressure time series ($\widetilde{p}_w = p_w/p^{\prime}_w$) at $Re_\tau \approx 14\,004$ for microphone $\mathcal{M}1$, over a short time interval, together with the magnitude of its Hilbert transform, $\vert H(\widetilde{p}_w) \vert$, and the de-meaned wall-pressure--squared time series. By visual inspection, these last two time series have similar large-scale energy content. Figure\,\ref{fig:hilbert}\textcolor{blue}{b} quantifies this further by overlaying the LCS for $u$ and $\vert H(\widetilde{p}_w)\vert$ and the LCS for $u$ and $p_w^2$ (those are identical to the ones shown in Fig.\,\ref{fig:up_coupling}\textcolor{blue}{b}). A remarkable collapse is observed for the two sets of LCS spectra, for all $Re_\tau$ cases considered. Hence, the large-scale variations in the wall-pressure intensity are directly linked to the passage of streamwise velocity fluctuations modulating the near-wall intensity \citep{tsuji:2015a}.

Gaining knowledge on how the coherence decays as a function of the streamwise separation between the velocity measurement and the wall-pressure sensor is highly relevant for real-time flow control (\emph{e.g.,} when sensors and actuators are separated to allow for control actions while the flow convects downstream). For our current data, the $\gamma^2_{up_w}$ and $\gamma^2_{up^2_w}$ coherence spectra are considered as a function of the streamwise distance of the velocity measurement (relative to the wall-pressure sensor at $x = 0$), for test case 3 corresponding to $Re_\tau \approx 14\,004$. Coherence spectra are shown in Figs.\,\ref{fig:up_coupling}\textcolor{blue}{c,d}, for $\gamma^2_{up_w}$ and $\gamma^2_{up^2_w}$, respectively. When increasing the streamwise distance, $\gamma^2_{up_w}$ decays with the coherence decreasing faster at smaller scales, as is expected. Similar conclusions were drawn for all other $Re_\tau$ test cases. When inspecting the decay in $\gamma_{up_w^2}^2$ (Fig.\,\ref{fig:up_coupling}\textcolor{blue}{d}), it becomes clear that the coherence with the quadratic wall-pressure term remains considerably larger than the one with the linear term. This means that the mechanism of large-scale modulation of the smaller-scale wall-pressure fluctuations (by the large-scale $u$ fluctuations) is dominant over the direct (linear) imprint of $u$ fluctuations on the wall-pressure.

\section{Coherence between wall-normal velocity and wall-pressure fluctuations}\label{sec:vwp}
\begin{figure} 
\vspace{0pt}
\centering
\includegraphics[width = 0.999\textwidth]{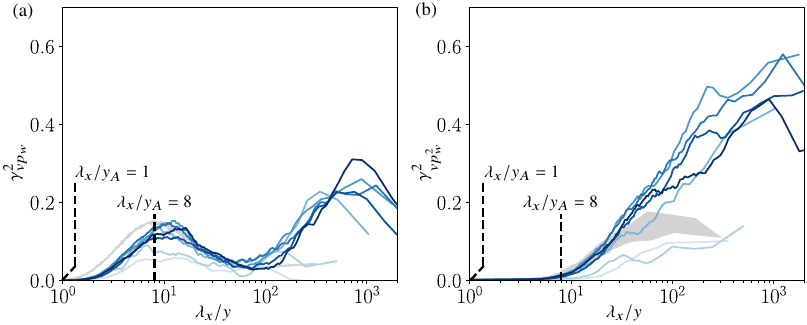}
\caption{(a) Coherence spectra for the fluctuations in wall-normal velocity and wall-pressure, and (b) the wall-normal velocity and wall-pressure--squared. Two sets of coherence spectra are shown, corresponding to velocity fluctuations measured at point $A$; an increase in colour intensity corresponds to an increase in $Re_\tau$ following test cases $1\rightarrow7$, listed in Table~\ref{tab:parameters}. The light grey shaded area is associated with the spread of coherence spectra from DNS data, as reported by \cite{baars:2024}. Note that all current coherence spectra are generated from temporal data, and plotted as spatial spectra by converting frequency into wavelength using $\lambda_x \equiv U_c/f$ with $U_c^+ = 10$.}
\label{fig:vwp}
\end{figure}

To further characterise the dynamics between fluctuations in velocity and wall-pressure, the foregoing presents a similar analysis as the one described in \S\,\ref{sec:up}, but instead of considering the $u$ fluctuations we concentrate on the $v$ fluctuations. Coherence spectra for $v$ and $p_w$ (Fig.\,\ref{fig:vwp}\textcolor{blue}{a}) exhibit local maxima around $\lambda_x/y \approx 10$ \citep[instead of $\lambda_x/y \approx 8$, as indicated by][]{baars:2024} and $\gamma_{vp_w}^2\approx 0.15$, a magnitude which is roughly 50\% higher than local maxima in the LCS for $u$ and $p_w$. Squaring the wall-pressure fluctuations brings higher levels of $\gamma_{vp_w^2}^2$ (see Fig.\,\ref{fig:vwp}\textcolor{blue}{b}), starting to rise at $\lambda_x/y \approx 3$. A similar Reynolds-number-independence is observed as was seen in the coherence analyses with the $u$ fluctuations.

Higher coherence between wall-normal velocity and wall-pressure fluctuations complies with earlier findings \citep[\emph{e.g.},][]{gibeau:2021}. In fact, given the non-permeability boundary condition, fluid motions directed towards the wall (with a negative $v$-component) will stagnate and thus give rise to a positive fluctuation in wall-pressure. Conversely, $v > 0$ fluctuations will tend to lower the unsteady wall-pressure. In general, the higher correlation between $v$ and $p_w$ can be ascribed to the Orr-mechanism \citep[see][among others]{jimenez:2013,luhar:2014}. Therefore, there exists a rather strong coupling between $v$ and $p_w$. Note, however, that our current work is motivated by using wall-pressure information to, eventually, predict the off-the-wall velocity fluctuations. Using pressure information to predict wall-normal velocity fluctuations for real-time control is not as effective as when considering streamwise velocity fluctuations, because the streamwise ones provide a stronger contribution to wall-shear stress generating mechanisms \citep[\emph{e.g.},][among others]{deck:2014}. Furthermore, $v$ fluctuations have a considerably shorter characteristic wavelength than $u$ fluctuations ($\lambda_x^+ \approx 250$ for the former, and $\lambda_x^+ \approx 1\,000$ for the latter). This not only requires faster processing for real-time operations, but would also constrain the sensor-actuator spacing due to the faster de-correlation of the $v$ fluctuations in the streamwise direction.

\section{Stochastic estimation of streamwise velocity fluctuations}\label{sec:est}
For real-time control purposes, wall-pressure sensing can be employed to generate an estimate of the fluctuating velocities in the logarithmic region of a wall-bounded turbulent flow. To this end, the foregoing will examine the accuracy of the prediction of $u$ fluctuations, performed with Linear and Quadratic Stochastic Estimation methods (LSE and QSE, respectively). These methods solely employ the time series of wall-pressure, and wall-pressure--squared, as the input quantities.

Estimates of the $u$ fluctuations in the logarithmic region at position $y_e$ can be formed through a convolution of time-domain kernels. Estimates of unconditional time series of off-the-wall velocity fluctuations, with the LSE and QSE procedures, follow from the formulations:
\begin{eqnarray}\label{eq:est}
   \widehat{u}_{\rm LSE}\left(y_e,t\right) &=& \left(h_l \circledast p_w \right)\left(t\right) \\
   \widehat{u}_{\rm QSE}\left(y_e,t\right) &=& \left(h_l \circledast p_w \right)\left(t\right) + \left(h_q \circledast p^2_w \right)\left(t\right),
\end{eqnarray}
where the (stochastic) temporal kernels of the linear term ($h_l$) and quadratic term ($h_q$) are the inverse Fourier transforms of the complex, frequency-domain kernels, \emph{e.g.}, $h_l\left(y_e,t\right) = \mathcal{F}^{-1}\left[H_L\left(y_e,f\right)\right]$ and similar for $h_q$. The linear kernel equals the cross-spectrum between $u$ and $p_w$, divided by the auto-spectra of $p_w$ (the input quantity during the estimation method),
\begin{equation}\label{eq:HL}
   H_L\left(y_e,f\right) = \frac{\langle \widetilde{U}\left(y_e,f\right) \widetilde{P}^*_w\left(f\right) \rangle}{\langle \vert \widetilde{P}_w\left(f\right) \vert^2 \rangle},
\end{equation}
whereas the quadratic kernel includes the wall-pressure--squared term, $p_w^2$, instead of the linear one,
\begin{equation}\label{eq:HQ}
   H_Q\left(y_e,f\right) = \frac{\langle \widetilde{U}\left(y_e,f\right) \widetilde{P}^*_{w,sq}\left(f\right) \rangle}{\langle \vert \widetilde{P}_{w,sq}\left(f\right) \vert^2 \rangle},
\end{equation}
with $\widetilde{P}_{w,sq}\left(f\right) = \mathcal{F}\left[p^2_w(t)\right]$. Given our current dataset, two estimation locations are considered ($y_e = y_A$ and $y_e = y_F$). Further details of the stochastic estimation procedures can be found elsewhere \citep{naguib:2001,baars:2024}.

\begin{figure} 
\vspace{0pt}
\centering
\includegraphics[width = 0.999\textwidth]{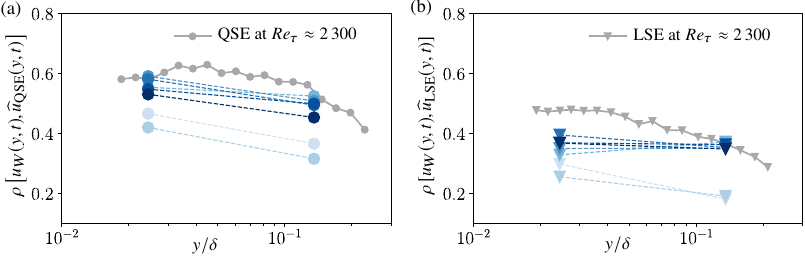}
\caption{(a) Correlation coefficient computed with the QSE-based streamwise velocity fluctuations in the logarithmic region, $\widehat{u}_{\rm QSE}$, and the reference time series, $u_W$. (b) Correlation coefficient computed with the LSE-based streamwise velocity fluctuations in the logarithmic region, $\widehat{u}_{\rm LSE}$, and the reference time series, $u_W$. Reference data are taken from \citet{baars:2024} at $Re_\tau \approx 2\,300$.}
\label{fig:corr_uhat}
\end{figure}

To evaluate the accuracy of the estimation with respect to the reference time series, $u(y_0,t)$, the Pearson correlation coefficient is employed. It is defined as the ratio of the covariance of two input signals to the product of the standard deviation of the two. Figure\,\ref{fig:corr_uhat}\textcolor{blue}{b} presents values of $\rho\left[u_W(y,t),\widehat{u}_{\rm QSE}(y,t)\right]= \mathrm{cov}\left[u_W(y,t),\widehat{u}_{\rm QSE}(y,t)\right]/(u_W' \widehat{u}'_{\rm QSE})$: the correlation coefficient between the reference time series $u_W(y,t)$ at points $A$ and $F$ (see Fig.\,\ref{fig:setup}\textcolor{blue}{b}) to the QSE-based time series, $\widehat{u}_{\rm QSE}(y,t)$. Here time series $u_W(y,t)$ is not equal to $u(y,t)$, because $u_W$ only retains wall-attached eddies. Effectively, $u_W$ is a large-scale pass-filtered signal of $u$, with its Reynolds number-invariant kernel characterised by a definitive cut-off at $\lambda_x/y = 14$ \citep{baars:2017a}.

With the exception of the data at the two lowest $Re_\tau$ test cases 1 and 2---whose time series are still affected by imperfect noise filtering (see \S\,\ref{sec:filtering})---collapse for all $Re_\tau$ test cases considered. A slight attenuation of $\rho$ is observed with respect to the reference data, which is attributed to a systematic error \citep{saccenti:2020} caused by additive facility noise. Furthermore, the data are in good agreement with the reference data from \cite{baars:2024} at $Re_\tau \approx 2\,300$. This result solidifies the conclusion of Reynolds-number independent estimation accuracy of wall-attached velocity fluctuations, based on wall-pressure sensing alone. As expected, a lower correlation coefficient appears in Fig.\,\ref{fig:corr_uhat}\textcolor{blue}{b}, where the estimation is performed with only the linear term of wall-pressure (thus using the LSE method). The performance of the statistical estimator now decays by about 20\% with respect to QSE-based estimation, across the given range of wall-normal distances.

\section{Conclusions}\label{sec:conclusion}
Statistical correlations of hydrodynamic wall-pressure and velocity fluctuations in the logarithmic region of a turbulent pipe flow were experimentally investigated. With a unique dataset acquired in the CICLoPE long-pipe facility, spanning a large range of friction Reynolds numbers ($4\,794 \lesssim Re_\tau \lesssim 47\,015$), this study reveals definitive Reynolds number trends of the scale-dependent wall-pressure--velocity coherence. For the linear coherence between the $u$ velocity (and $v$ velocity) and wall-pressure, a Reynolds-number-independent scaling of the coherence spectra appears at the intermediate scale range when scaled with distance-from-the-wall. This trend is also statistically similar across several wall-bounded flows when compared to the data available from the open literature. When the squared wall-pressure fluctuations are considered instead of the linear wall-pressure, the coherence spectra for the wall-pressure and velocity fluctuations are higher in amplitude at the (very) large-scale end of the spectra. Physically, this link between wall-pressure--squared and velocity typifies a modulation effect as squaring the wall-pressure introduces low-frequency content that is reflective of how the higher-frequency wall-pressure intensity varies. Current findings of the coherence spectra bear relevance to stochastic estimation schemes, in which wall-pressure can be considered as an input to estimate off-the-wall velocity fluctuations. With the aid of a quadratic stochastic estimation method, it was shown that for each $Re_\tau$ investigated the estimated time series and at rue temporal measurement of velocity inside the turbulent pipe flow yielded a normalized correlation coefficient of up to $\rho \approx 0.6$ (while this was below 0.4 for a linear stochastic estimation method excluding the wall-pressure--squared term). This demonstrates that (sparse) wall-pressure sensing can be employed for meaningful estimation of off-the-wall velocity fluctuations. And, that wall-pressure as an input for estimation schemes is scalable to application-level conditions.

\vspace{15pt}
\backsection[Acknowledgements]{We thank the Dipartimento di Ingegneria Industriale (DIN Department) of the University of Bologna for granting a Young Visiting Fellowship, enabling an on-site visit for GD and WJB. We also acknowledge the technical staff in Forl\`{i} for the support during the experiments.}

\backsection[Funding]{The work of the UniBo team was carried out within the MOST – Sustainable Mobility National Research Center and received funding from the European Union Next-GenerationEU [PIANO NAZIONALE DI RIPRESA E RESILIENZA (PNRR) – MISSIONE 4 COMPONENTE 2, INVESTIMENTO 1.4 – D.D. 1033 17/06/2022, CN00000023].}

\backsection[Declaration of interests]{The authors report no conflict of interest.}

\backsection[Author ORCIDs]{\\Giulio Dacome \hyperlink{https://orcid.org/0009-0000-3088-2495}{https://orcid.org/0009-0000-3088-2495}; \\Lorenzo Lazzarini 
\hyperlink{https://orcid.org/0000-0002-3951-5242}{https://orcid.org/0000-0002-3951-5242}; \\Alessandro Talamelli
\hyperlink{https://orcid.org/0000-0002-5999-8990}{https://orcid.org/0000-0002-5999-8990}; \\Gabriele Bellani 
\hyperlink{https://orcid.org/0000-0001-8502-1991}{https://orcid.org/0000-0001-8502-1991}; \\Woutijn J. Baars \hyperlink{https://orcid.org/0000-0003-1526-3084}{https://orcid.org/0000-0003-1526-3084}.}


\appendix
\section{Removing facility noise from the experimental wall-pressure signals}\label{app:noise}
Wall-pressure measurements by means of microphones, mounted within sub-surface-cavities communicating with the flow through a pinhole orifice, result in signal contamination from two main sources: (1) acoustic noise from the flow facility, and (2) acoustic resonance as a consequence of the pinhole--sub-surface-cavity geometry. While a correction for the latter can directly be implemented in the frequency domain and takes the form of a division of the spectrum by the gain-squared of a correction kernel (as done in \S\,\ref{sec:pressure}), the former requires a more elaborate procedure. In particular, when considering a raw pressure time series of one of the microphones in Fig.\,\ref{fig:setup}\textcolor{blue}{c}, it is necessary to disambiguate hydrodynamic wall-pressure signatures from the one induced by acoustic phenomena. In the case of turbulence-induced fluctuations, especially wall-pressure, they possess negligible streamwise and spanwise (azimuthal) coherence when considering relatively large sensor separations. Acoustic pressure fluctuations, however, convect from sensor to sensor retaining high correlation between detection stations, both in the streamwise and spanwise directions directions of the flow.

With the experimental setup illustrated in \S\,\ref{sec:methods}, the acoustic waves produced by the operation of the CICLoPE facility will be detected by all microphones embedded with the aid of the pinhole--sub-surface-cavity, $\mathcal{M}1$ to $\mathcal{M}4$, and the microphone mounted along the centreline of the pipe, $\mathcal{M}5$. However, the pressure time series measured by $\mathcal{M}5$ will not contain hydrodynamic wall-pressure fluctuations. Removing facility noise requires the identification of spatial modes that are correlated among the spatially-separated sensors and whose signatures are also detected by the centreline microphone. Harmonic proper orthogonal decomposition (hPOD) is suitable to identity these modes as, compared to conventional POD, the spatial decomposition is performed in the spectral domain, which accounts for phase-shifts of pressure signatures between sensors \citep[see][]{tinney:2020}. During hPOD, a signal is decomposed into complex-valued and frequency-dependent eigenvalues and eigenmodes. These eigenvalues and eigenmodes follow from solving an eigenvalue problem with the harmonic complex-valued kernel. This kernel, denoted as $\check{R}$, contains the spectral cross-correlation of all possible combinations of two pressure signals, with entries of the matrix being constructed according to:
\begin{equation}\label{eq:sPOD_kernel}
    \check{R}_{ij}\left(\mathbf{x},\mathbf{x}';f\right) = \langle \widetilde{P}_{w,r;i}\left(\mathbf{x};f\right) \widetilde{P}^*_{w,r;j}\left(\mathbf{x};f\right) \rangle,
\end{equation}
with $\widetilde{P}_{w,r;i}\left(\mathbf{x};f\right) = \mathcal{F}\left[p_{w,r;i}(\mathbf{x},t)\right]$, and subscripts $i$ and $j$ denoting the different time series of the raw measured pressure. Position vector $\mathbf{x}$ contains the sensor coordinates (\emph{e.g.}, $\mathbf{x}_{\mathcal{M}1}$ is the position vector of microphone $\mathcal{M}1$). Spectral eigenvalues are denoted as $\Lambda^{(n)}(f)$; for each mode number $n$ (a total of $n = 1 \dots N_m$ modes, with $N_m = 5$ being equal to the number of sensors) this frequency-dependent eigenvalue has $N_f$ entries. Here $N_f$ is the temporal FFT ensemble size considered ($N_f = 2^{15}$, resulting in a frequency resolution of d$f = 1.56$\,Hz). Harmonic eigenmodes are space- and frequency-dependent and denoted as $\Phi^{(n)}\left(\mathbf{x},f\right)$; for each mode number $n$ the modes have dimensions of $N_m \times N_f$. Finally, the original pressure signal can be reconstructed (in the frequency domain) using the summation of all modes,
\begin{equation}
    \tilde{P}_{w,r}(\mathbf{x};f) = \sum_{n}{\check{A}^{(n)}\left(f\right)\Phi^{\left(n\right)}\left(\mathbf{x};f\right)},
\end{equation}
with $\check{A}^{(n)}(f)=\int \widetilde{P}_{w,r}(\mathbf{x};f)\Phi^{(n)}(\mathbf{x};f)\mathrm{d}\mathbf{x}$ being the frequency-dependent complex random expansion coefficients.

\begin{figure} 
\vspace{0pt}
\centering
\includegraphics[width = 0.999\textwidth]{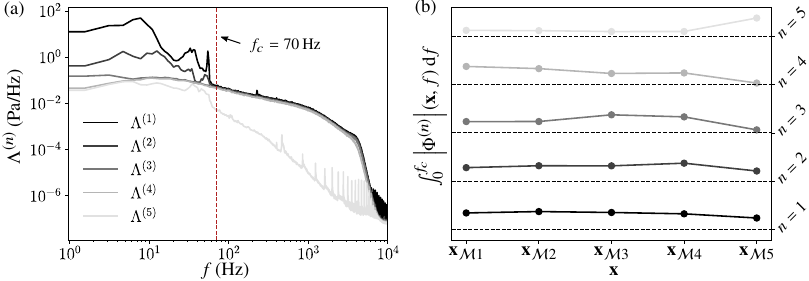}
\caption{(a) Spectra of the eigenvalue obtained from the complex-valued $\check{R}$ kernel, for test case 3 ($Re_\tau \approx 14\,004$). (b) Normalised magnitude of the complex modes $\Phi^{(n)}$, for $n = 1 \dots 5$, integrated over the range $0 < f \lesssim 70$\,Hz. Each curve is offset by one unit vertically for graphical readability.}
\label{fig:hpod}
\end{figure}

For one of the current datasets (test case 3, $Re_\tau \approx 14\,004$), the five frequency-dependent eigenvalues are shown in Fig.\,\ref{fig:hpod}\textcolor{blue}{a}. The first two eigenvalues contain clear signatures of facility noise, especially at the low frequencies. The spectra of the first four eigenvalues show a broadband distribution in the mid-to-high frequency band, whereas the fifth eigenvalue only has significant energy in the low-frequency band at $f \lesssim 60$\,Hz. To determine which mode set to retain for filtering the wall-pressure time series, the spatial distribution of eigenmodes is also examined. In particular, by construction of the experiment, the ideal set of modes to retain consists of the ones that exhibit no activity at the centreline microphone, $\mathcal{M}5$. To aid in the selection of modes, we only consider frequencies in the range $0 < f < f_c$, with $f_c = 70$\,Hz, as the facility noise is concentrated in this band. The magnitude of the eigenmodes, integrated over the aforementioned frequency range, is displayed in Fig.\,\ref{fig:hpod}\textcolor{blue}{b}. Upon inspection of the five curves, it is clear that modes 3 and 4 are the ones encompassing negligible activity at the position of the centreline microphone, $\mathbf{x}_{\mathcal{M}5}$. Based on this, it was decided to reconstruct the wall-pressure time series with modes 3 and 4 only. And so, the filtered wall-pressure time series can be computed as the inverse Fourier transform of the frequency-dependent lower-order:
\begin{equation}
    \widetilde{P}_{w,f}\left(\mathbf{x};f\right) = \sum_{n=\{3,4\}}{\check{A}^{\left(n\right)}\left(f\right)\Phi^{\left(n\right)}\left(\mathbf{x};f\right)}~~~\rightarrow~~~p_{w,f}\left(x,t\right) = \mathcal{F}^{-1}\left[\widetilde{P}_{w,f}(\mathbf{x};f)\right].
\end{equation}

For the other friction Reynolds numbers considered in this study (see Table\,\ref{tab:parameters}), a similar procedure was applied. Similar conclusions could be drawn in regards to the selection of modes to retain for filtering, with the only minor difference lying in the selection of the upper frequency bound for acoustic contamination, $f_c$. For increasing Reynolds numbers, $f_c$ increases; physically this is caused by a larger blade passing frequency of the axial fans operating the pipe flow facility.

\bibliographystyle{jfm}
\bibliography{bibtex_database}

\end{document}